\documentclass{bmvc2k}
\usepackage{float}
\usepackage{bm}
\usepackage{array}
\usepackage{booktabs}
\usepackage{wrapfig}
\usepackage{graphicx}
\usepackage{epsfig}
\usepackage{epstopdf}
\usepackage{amsmath}
\usepackage{amssymb}
\usepackage{multirow}
\usepackage{booktabs}
\usepackage{algorithm}
\usepackage{algorithmic}
\usepackage{bbding}
\title{Revisiting Temporal Modeling for Video Super-resolution}

\addauthor{Takashi Isobe}{jbj18@mails.tsinghua.edu.cn}{1}
\addauthor{Fang Zhu}{zhuf14@gmail.com}{2}
\addauthor{Xu Jia}{x.jia@huawei.com}{3}
\addauthor{Shengjin Wang}{wgsgj@tsinghua.edu.cn}{1}

\addinstitution{
Department of Electronic Engineering, \\
Tsinghua University
}
\addinstitution{
Department of Computer Engineering, \\
New York University, \\
11201, USA
}
\addinstitution{
Noah's Ark Lab, \\ Huawei Technologies
}

\runninghead{T. ISOBE}{Revisiting Temporal Modeling for Video Super-resolution}

\begin{document}
	\maketitle
	
	\begin{abstract}
		
		Video super-resolution plays an important role in surveillance video analysis and ultra-high-definition video display, which has drawn much attention in both the research and industrial communities. Although many deep learning-based VSR methods have been proposed, it is hard to directly compare these methods since the different loss functions and training datasets have a significant impact on the super-resolution results. In this work, we carefully study and compare three temporal modeling methods (2D CNN with early fusion, 3D CNN with slow fusion and Recurrent Neural Network) for video super-resolution. We also propose a novel Recurrent Residual Network (RRN) for efficient video super-resolution, where residual learning is utilized to stabilize the training of RNN and meanwhile to boost the super-resolution performance. Extensive experiments show that the proposed RRN is highly computational efficiency and produces temporal consistent VSR results with finer details than other temporal modeling methods. Besides, the proposed method achieves state-of-the-art results on several widely used benchmarks. Code is available at \url{https://github.com/junpan19/RRN}.
	
	\end{abstract}
	%-------------------------------------------------------------------------

	\section{introduction}
	Super-resolution is a traditional yet still dynamic topic in low-level vision field, which aims at producing a high-resolution image from the corresponding low-resolution counterparts. It has drawn much attention in recent years due to the increasing demand in mobile phones and ultra-high-definition displays.
	Single-image super-resolution (SISR) has achieved significant improvements over the last few years, which benefits greatly from progress in deep learning. Recently, more attentions have been shifted to the video super-resolution (VSR) since a video sequence should contain more abundant information. In contrast with SISR, which relies on natural image priors and self-similarity within images for recovering the missing details, VSR is able to utilize additional temporal information from the neighboring frames for further improving the quality of SR.
	\\
	Previous VSR works divide into two categories: 1) explicit motion compensation based methods~\cite{kappeler2016video,caballero2017real,tao2017detail,sajjadi2018frame,xue2019video} and 2) implicit motion compensation based methods~\cite{kim20183dsrnet,jo2018deep,haris2019recurrent,wang2019edvr,Fuoli-arxiv19-rlsp,yi2019progressive}. As for explicit motion compensation based methods, kappeler~\textit{et at.} proposes to warp all neighboring frames to the reference frame based on the offline estimated optical flow; VESCPN~\cite{caballero2017real} is the first end-to-end video SR method by jointly training optical flow estimation and spatial-temporal networks. However, these works are not ideal for VSR since inaccurate motion estimation and alignment would result in errors and deteriorated super-resolution performance. Besides, the computation of optical flow often introduces heavy computational load, which restricts deploying these methods in real systems. 
	\\
	Another branch of VSR explores advanced temporal modeling frameworks to utilize motion information in an implicit manner. Typically, there temporal modeling framework have been widely used: 2D with early fusion CNN~\cite{tian2018tdan,haris2019recurrent,wang2019edvr,yi2019progressive}, 3D CNN with slow fusion~\cite{kim20183dsrnet,jo2018deep,isobe2020videocvpr} and Recurrent Neural Network (RNN)~\cite{huang2015bidirectional,sajjadi2018frame,Fuoli-arxiv19-rlsp,isobe2020videoeccv}. 

	Although extensive experiments have been reported on the aforementioned methods, it is hard to directly compare the effectiveness of these temporal modeling approaches because they adopt different training sets and loss functions to develop their model, which significantly influences the quality of the estimated high-resolution frames. For example,~\cite{jo2018deep} trained their model on the private dataset and supervised an elaborately designed Huber Loss.~\cite{haris2019recurrent} developed their model on Vimeo-90k~\cite{xue2019video} dataset and supervised by $\mathcal{L}1$ Loss. Moreover, different network depth also limits the direct comparison among these temporal modeling methods, \textit{e,g.}~\cite{jo2018deep} adopted 52 layers in their large model and~\cite{wang2019edvr} exploited more deep network. 
	\\

	In this paper, we comprehensively investigate the effectiveness of different temporal modeling approaches on the VSR task by using the fixed loss function ($\mathcal{L}1$ Loss) and training data. Specifically, we explore three commonly used temporal modeling methods: 1) 2D CNN with early fusion, 2) 3D CNN with slow fusion and 3) RNN. Inspired by~\cite{lim2017enhanced}, we design the 2D CNN with several modified 2D residual blocks. As for 3D CNN, we further modify these 2D residual blocks to 3D residual blocks. We also incorporate such residual connection into the hidden state of the recurrent network and propose Recurrent Residual Network (RRN) for video super-resolution. In the proposed hidden state, the identity branch not only carries rich image details from the previous layers to the next layers but also helps to avoid gradient vanishing in RNN training. For fair comparison of these temporal modeling methods, we evaluate these models on widely used benchmarks with the same network depth. The experimental results show that Recurrent-based methods are highly efficient and effective in dealing with the VSR problem. Besides, the proposed RRN achieves state-of-the-art performance on three benchmarks. 

	To sum up, we make the following contributions:
	\begin{itemize}
		\item
		We carefully study three commonly used temporal modeling methods (2D CNN with early fusion, 3D CNN with slow fusion, and RNN) for the VSR problem.
		\item
		We propose a novel hidden state for the recurrent network, which achieves the best performance among all temporal modeling methods. To more surprise, the proposed method outperforms the previous state-of-the-art methods on all three public benchmarks.

	\end{itemize}
	
	\section{Related work}
	\begin{figure}[t]
		\includegraphics[width=1\textwidth] {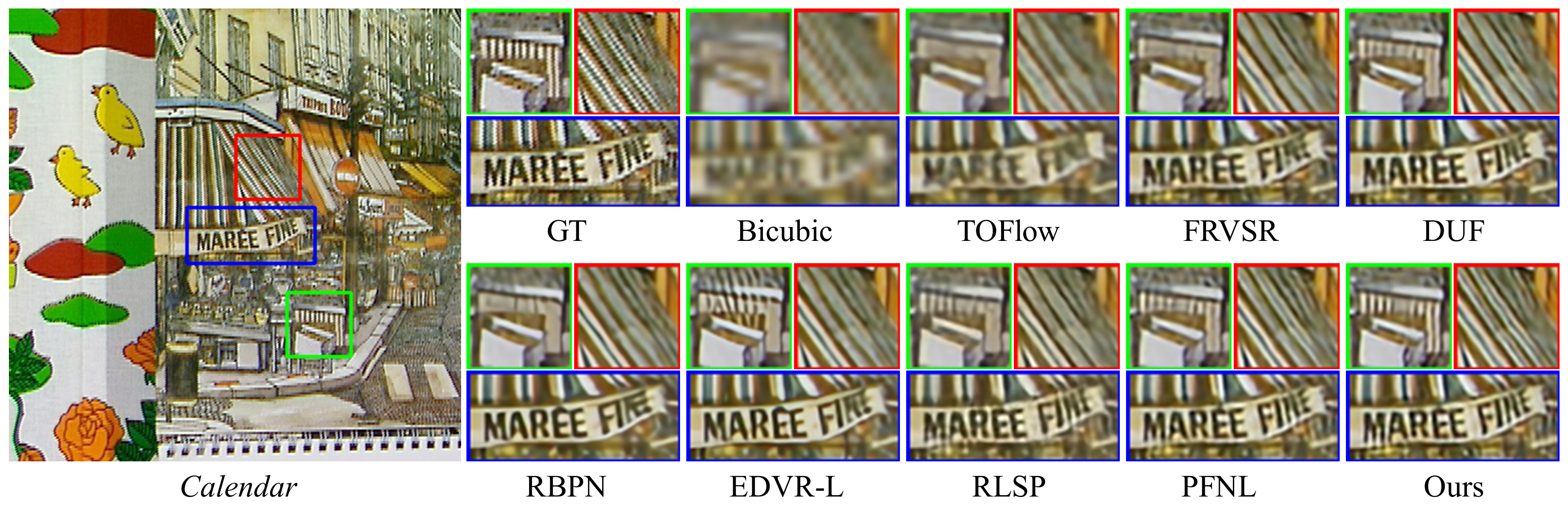}
		\vspace{-5mm}
		\caption{\small $\times$ 4 VSR results for the scene \textit{Calendar} in Vid4~\cite{liu2013bayesian} dataset. Our method produces sharper edges and more detailed textures than other state-of-the-art methods.}
		\vspace{-3mm}
		\label{fig1} 
	\end{figure}
	\textbf{Single image super-resolution.}
	With the development of deep learning, the convolutional neural network method has achieved dramatic advantages against conventional methods in single image super-resolution (SISR). In~\cite{dong2014learning}, Deng~\textit{et at.} first proposed to use a 3-layer end-to-end convolutional neural network to fill the missing details of the interpolated low-resolution images and showed promising results. Since then, there have been numerous learning-based SISR methods~\cite{kim2016accurate,kim2016deeply,tai2017image,zhang2018residual,haris2018deep,zhang2018image} constantly emerging. VDSR~\cite{kim2016accurate} further improved CNN depth by stacking more convolutional layers with global residual learning. DRCN~\cite{kim2016deeply} first proposed to reduce model parameters with recursive learning in deep CNN. However, DRCN also suffers from performance degradation problems when increasing more convolutional layers (up to 16 convolutional recursions). To further increase CNN depth, DRRN~\cite{tai2017image} was proposed with local residual learning and global residual learning strategy. Much deeper CNN, including~RDN~\cite{zhang2018residual},~DBPN~\cite{haris2018deep},~RCAN~\cite{zhang2018image} were then introduced, which outperformed previous works by a large margin. In this paper, we also enjoy the merits of residual learning and incorporate residual connection in the Recurrent neural network (RNN). The proposed RRN not only carries rich details from the previous layers to later layers in the hidden state, where information can be stably propagated even through a large number of convolutional layers but also carries on historical information through a long range of time steps as the additional complementary information for the later time step. \\
	\textbf{Video super-resolution.}
	Temporal modeling plays a key role in VSR. Previous works performing temporal aggregation fall into three branches: 1) 2D CNN with early fusion~\cite{tian2018tdan,haris2019recurrent,wang2019edvr,yi2019progressive} 2) 3D CNN with slow fusion~\cite{kim20183dsrnet,jo2018deep} and 3) RNN based~\cite{sajjadi2018frame,Fuoli-arxiv19-rlsp} methods. TDAN~\cite{tian2018tdan} and EDVR~\cite{wang2019edvr} aggregated multi-frames features with several 2D convolutional layers on the top of the feature-wise alignment. PFNL~\cite{yi2019progressive}  captured long-range dependencies through a kind of non-local operation, and then aggregated the correlations maps with several 2D convolutional layers.Kim~\textit{et at.} used several stacked 3D convolutional layers to extract both spatial and temporal information within a temporal sliding window in a slow fusion manner and implemented this fashion over the entire video sequence. DUF~\cite{jo2018deep} estimated dynamic filters with stacked 3D convolutional layers for implicit motion compensation and upsampling. With the advanced temporal modeling strategy, CNN based methods show superior performance on several benchmarks. However, the overlap of sliding windows leads to redundant computation, which limits the VSR efficiency. As for recurrent based methods, both historical information across time step and current information between consecutive frames can be used to enhance details for an LR frame. In~\cite{sajjadi2018frame}, Sajjadi~\textit{et at.} proposed to conduct motion estimation and warp operation between the previous frame and current frame, and then super-resolve the aligned frame in a recurrent manner. However, inaccurate motion estimation would cause severe artifacts and increase the risk of error accumulation. Recently, Fouli~\textit{et at.} proposed RLSP~\cite{Fuoli-arxiv19-rlsp}, which propagated historical information in feature space and without explicit motion estimation. Related to \cite{Fuoli-arxiv19-rlsp}, our method also propagates historical information in feature space. However, in~\cite{Fuoli-arxiv19-rlsp}, they adopt seven simply connected convolutional layers as the hidden state, which is difficult to preserve texture details when propagate so many layers in the hidden state. In addition, RLSP fed three consecutive frames into each hidden state. With more input frames, the hidden state would easily suffers from error accumulation, especially when there is large motion between consecutive frames. In this work, we exploit two frames (previous and current) as hidden state input, and incorporate identity mapping in hidden state to preserve the texture details through so many layers.

	\section{Methodology}
	\subsection{Overview}
\begin{figure}[t]
		\centering
		\includegraphics[width=1\textwidth] {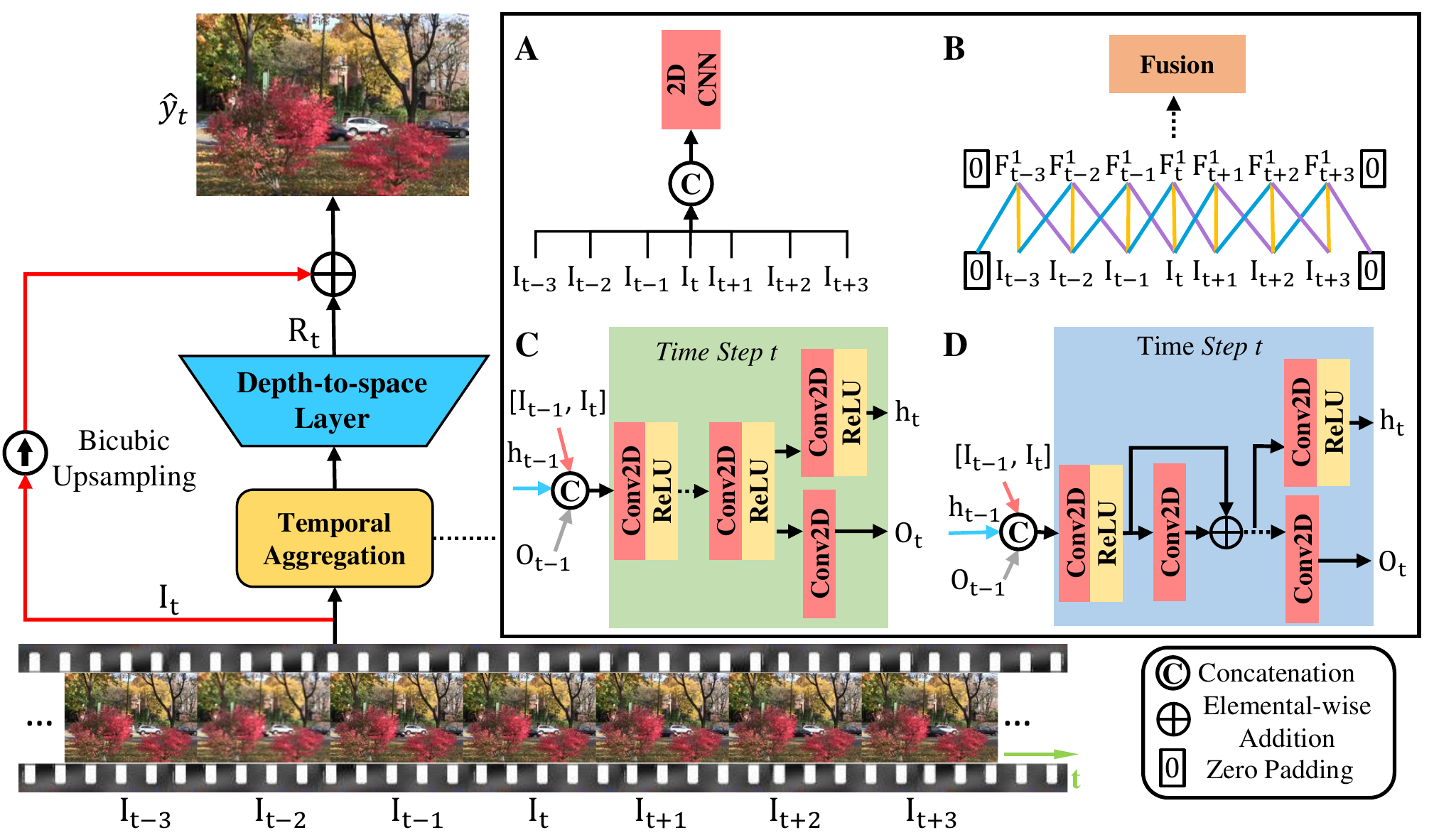}
		\vspace{-5mm}
		\caption{\small Schematic illustration of three commonly used temporal modeling frameworks (A: 2D CNN with early fusion, B: 3D CNN with slow fusion, C: RNN). D is the proposed RRN.}
		\label{pipeline} 
		\vspace{-5mm}
	\end{figure}

	In this section, we introduce the overall system pipeline and detailed configurations of the temporal modeling methods. The whole system consists of two parts: a temporal modeling network which takes consecutive frames as input and integrates them with the reference frame, and a loss function to optimize the network utilizing motion information in an implicit manner. We comprehensively study and compare three temporal modeling methods, including 2D CNN with early fusion, 3D CNN with slow fusion and RNN. Schematic illustration of these networks is shown in Fig.~\ref{pipeline} (a), (b) and (c), respectively. The detailed architecture about the proposed hidden state is shown in Fig.~\ref{pipeline} (d). 
	\subsection{Network Design}
	We consider three types of deep neural networks: (1) 2D CNN (2) 3D CNN and (3) RNN. For 2D CNN, the input frames are first concatenated along the channel axis, and then aggregated with a stack of 2D convolutional layer. 3D CNN takes a video sequence as input and then exploits a stack of 3D convolutional layer to extract the spatial-temporal information from the video sequence. Comparing to the CNN methods, RNN takes fewer frame as the hidden state input and handles a long video sequence with a recurrent manner.
	
	\textbf{2D CNN with early fusion.}

	Inspired by~\cite{lim2017enhanced}, we design the 2D CNN with several modified 2D residual blocks, where each block consists of a $3\times 3$ convolutional layer followed by ReLU~\cite{glorot2011deep}. Such model takes $2T+1$ consecutive frames as input. The aggregation process can be formulated as:
	\begin{equation}
	R_t = W_{net2D}\{W_{fusion2D}[I_{t-T},\dots,I_{t+T}]\}
	\end{equation}
	Where $[\cdot,\cdot,\cdot]$ denotes the concatenation operation. The input tensor shape of $W_{fusion2D}$ is $NC\times H \times W$, where $N=2T+1$, $C$ is the number of color channels, and $H$, $W$ are the height and weight of the input LR frames, respectively. $W_{fusion2D}$ and $W_{net2D}$ represent a set of weights (the biases are omitted to simplify notations) of the early fusion layer and 2D CNN, respectively. The shape of the produced residual map is $R_t$ is $H\times W \times Cr^2$, where r is the upscale factor. The high-resolution residual maps $R^\uparrow_t$ is obtained by adopting depth-to-space operation~\cite{shi2016real}. Finally, the high-resolution image $\hat{y}_t$ is obtained by adding the predicted high-resolution residual map $R^\uparrow_t$ to a bicubic up-sampled high-resolution reference image $I^\uparrow_t$ at the end of the network.
	
	\textbf{3D CNN with slow fusion.} In 3D CNN, we modify the 2D convolutional layers in the 2D residual blocks to $3\times3\times3$ convolutional layers for extracting spatial-temporal information. We use the same network depth for both 2D and 3D CNN for a fair comparison. In contrast with 2D CNN, 3D CNN takes a video sequence as input and extracts the spatial-temporal information in a slow fusion manner. Specifically, a 3-dimensional filter moves both in temporal and spatial axis directions to extract both spatial and temporal information. Typically, the temporal dimension depth of 3D filter is much smaller than the length of the input sequence. Such slow fusion process can be described as:
	\begin{equation}
	R_t = W_{fusion3D}\{W_{net3D}\{I_{t-T:t+T}\}\}
	\end{equation}
	Where $W_{net3D}$ and $W_{fusion3D}$ represent a set of weights (the biases are omitted to simplify notations) of 3D CNN and the later fusion layer, respectively. The input tensor shape of $W_{net3D}$ is $C\times N \times H \times W$, where $C$ is the number of color channels, $N=2T+1$, and $H$, $W$ are the height and weight of the input LR frames, respectively. To prevent the number of frames from decreasing, we add two frames with pixel value of zero in the temporal axis.
	% the temporal information integration process from other frames to the reference frame is conducted without explicitly taking the reference frame into consideration.
	
	\textbf{RNN.} Typically, a hidden state at time step $t$ takes three parts as input: (1) the previous output $o_{t-1}$, (2) the previous hidden state features $h_{t-1}$ and (3) two adjacent frames $I_{\{t-1,t\}}$. Intuitively, in an video sequence, pixels within successive frames usually bear a strong similarity. The high-frequency texture details in $t$-th time step should be further refined by borrowing the complementary information from the previous layer. However, RNN in VSR~\cite{Fuoli-arxiv19-rlsp} also suffers gradient vanishing issue as many other video processing tasks~\cite{chen2019temporal,zhang2019eleatt,gordon2018re}. To address this issue, we propose a novel recurrent network, termed as Residual Recurrent Network (RRN), which adopts residual mapping between layers with identity skip connections. Such design ensures a fluent information flow and has the ability to preserve the texture information over long periods making RNN easier to process a longer sequences, and meanwhile reduce the risk of gradient vanishing in training. At time step $t$, the RRN uses following equations to generate output $h_t$ and $o_t$ for the next time step $t$+1:
	\begin{equation}
	\begin{aligned}[a]
	& \hat{x}_0 = \sigma(W_{conv2D}\{[I_{t-1},I_{t},o_{t-1},h_{t-1}]\}) &&\hat{x}_k=g(\hat{x}_{k-1}) + \mathcal {F}(\hat{x}_{k-1}), k\in[1,K] \\
	&  h_t = \sigma(W_{conv2D}\{\hat{x}_K\})  &&  o_t=W_{conv2D}\{\hat{x}_K\} 
	\end{aligned}
	\end{equation}
	\\

	Where $\sigma$($\cdot$) represents the ReLU function. $g(\hat{x}_{k-1})$ denotes an identity mapping in $k$-th residual block: $g(\hat{x}_{k-1})=\hat{x}_{k-1}$, and $\mathcal {F}(\hat{x}_{k-1})$ denotes the residual mapping to be learned.

	\section{Experiment}
	\subsection{Dataset}
	Previous works use different training sets and different down-sampling kernels, which restricts fair comparisons. In this work, we adopt Vimeo-90k~\cite{xue2019video} as the training set. Vimeo-90k is a public dataset for video restoration tasks, including video denoising, deblocking as well as super-resolution. Vimeo-90k contains around 90k 7-frame video clips with various motions and diverse scenes. To develop our model, the low-resolution patches in the size of $64\times 64$ are obtained by applying Gaussian blur with $\sigma=1.6$ to a high-resolution frame and further downsampling by $4\times$ scale factor. We evaluate the developed models on Vid4~\cite{liu2013bayesian}, SPMCS~\cite{tao2017detail} and UDM10~\cite{yi2019progressive} datasets. Vid4 consists of four scenes with various motion and occlusion. SPMCS and UDM10 are the recently proposed validation sets, which contain diverse senses with considerable high-resolution frames than Vid4.

	\begin{figure}[t]
		\includegraphics[width=1\textwidth] {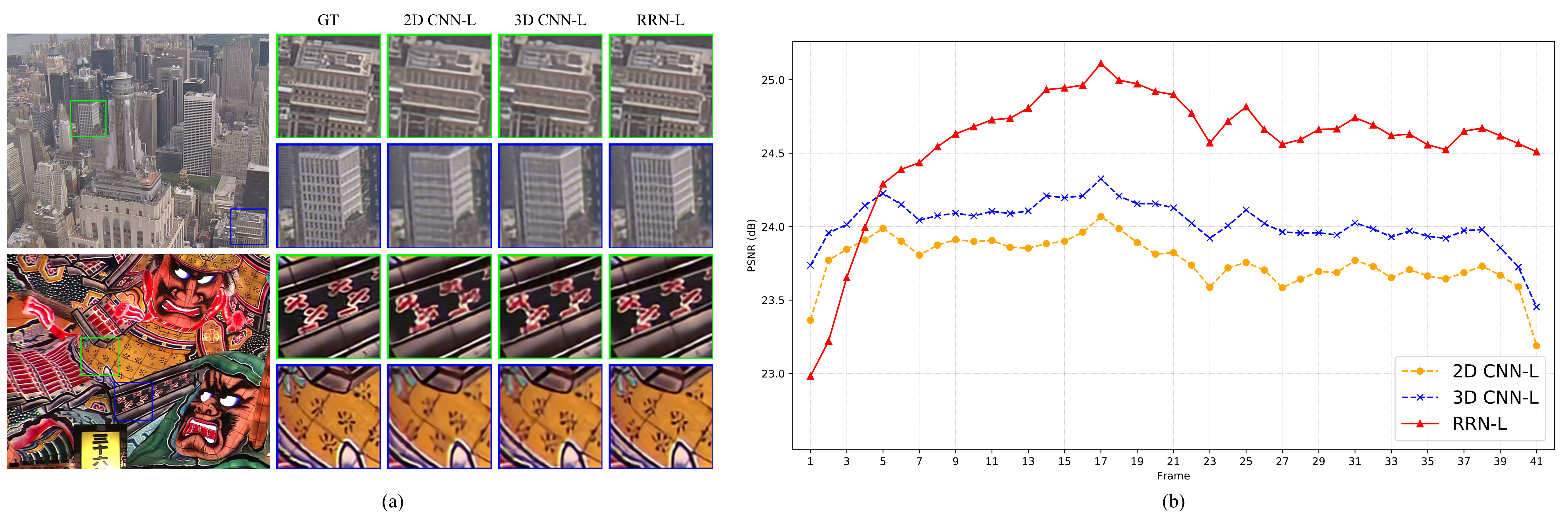}
		\vspace{-5mm}
		\caption{\small (a): Qualitative results for different temporal modeling methods on \textbf{Vid4}~\cite{liu2013bayesian} and \textbf{SPMCS}~\cite{tao2017detail} dataset for 4$\times$VSR. (b): Information flow over time for 2D CNN, 3D CNN and RNN on the \textit{Calendar} sequence.}
		\label{modeling} 
		\vspace{-3mm}
	\end{figure}

	\begin{table}[t]
		\begin{center}
			\scalebox{0.8}
			{	
				\begin{tabular}{|l||c|c|c||c|c|c|}
					\hline
					Method    & \begin{tabular}{@{}c@{}} 2D CNN\\ S \end{tabular}  & \begin{tabular}{@{}c@{}} 3D CNN\\ S \end{tabular}  & \begin{tabular}{@{}c@{}} RRN\\ S \end{tabular}  & \begin{tabular}{@{}c@{}} 2D CNN\\ L  \end{tabular}   & \begin{tabular}{@{}c@{}} 3D CNN\\ L \end{tabular} &\begin{tabular}{@{}c@{}} RRN\\ L \end{tabular} \\ \hline \hline
					Blocks     &5     &5       &5       &10	    &10  &10   \\	\hline
					Input Frames  &7     &7   &recurrent     &7       &7  &recurrent \\	\hline
					\# Param. [M]  &2.8   &5.5   &1.9    &4.3   &9.9    &3.4 \\	\hline
					FLOPs [GMAC]   &395   &1852  &108   &480    &3640  &193 \\ \hline \hline
					Vid4 (Y)   &26.72  &27.08   &{\color{blue}27.38}   &26.96
					&27.25&{\color{red} 27.69 }  \\	\hline
					SPMCS (Y)   &29.05  &29.42   &29.48  &29.51  &{\color{blue}29.64}  &{\color{red}29.84} \\	\hline
					UDM10 (Y)   &37.67  &38.12  &38.33  &38.15 &{\color{blue}38.43}&{\color{red}38.97} \\	\hline
					Runtime [ms]     &97   &558    &30     &116     &1045    &45  \\ \hline
				\end{tabular}
			}
		\end{center}
	
		\caption{\small Comparison of PSNR values on \textbf{Vid4~\cite{liu2013bayesian}, SPMCS~\cite{tao2017detail} and UDM10~\cite{yi2019progressive}} and runtime between different temporal modeling methods for $\times$ 4 VSR. Y denotes the evaluation on luminance channel. Runtime is calculated on an LR image of size 320$\times$180. {\color{red}Red} text indicates the best and {\color{blue} blue} text indicates the second best performance. Best view in color.}
		\label{model_tab}
		\vspace{-5mm}
	\end{table}
	
	\subsection{Implementation Details}
	We consider two models which have different network depth for all temporal modeling method.

	As for 2D CNN, 2D CNN-S, 2D CNN-L adopt five and ten 2D residual blocks, respectively. As for 3D CNN, 3D CNN-S and 3D CNN-L adopt five and ten 3D residual blocks, respectively. 

	The channel size for 2D CNN and 3D CNN is set to 128. For a fair comparison with the CNN based methods, we also adopt five and ten residual blocks as the hidden state for RRN-S and RRN-L, respectively. Each block consists of a convolutional layer, a ReLU layer and following another convolutional layer. The channel size of convolutional layer is set to 128. At the time step $t_0$, the previous estimation is initialized with zero. To train the CNN based models, the learning rate is initially set to $1\times10^{-4}$ and $1\times10^{-3}$ for 2D CNN and 3D CNN, respectively, and multiplied by 0.1 after 10 epochs. The training step completes after 30 epochs. To train the RNN based model, the learning rate is initially set to $1\times10^{-4}$ and later down-scaled by a factor of 0.1 every 60 epoch till 70 epochs. All models are supervised by pixel-wise $\mathcal{L}1$ loss function with Adam~\cite{kingma2014adam} optimizer by setting $\beta_1 = 0.9$, $\beta_2 = 0.999$ and weight decay of $5\times 10^{-4}$. We set the size of mini-batch as $64$ and $4$ for CNN based and RNN based methods, respectively. The $\mathcal{L}1$ loss is applied on all pixels between the ground truth frames $y^\star_t$ and the network’s output $\hat{y}_t$, defined by $\mathcal{L} = \Vert y^\star_t - \hat{y}_t \Vert$. All experiments are conducted using Python 3.6.4 and Pytorch 1.1.

	\begin{figure*}[h]
		\includegraphics[width=1\textwidth] {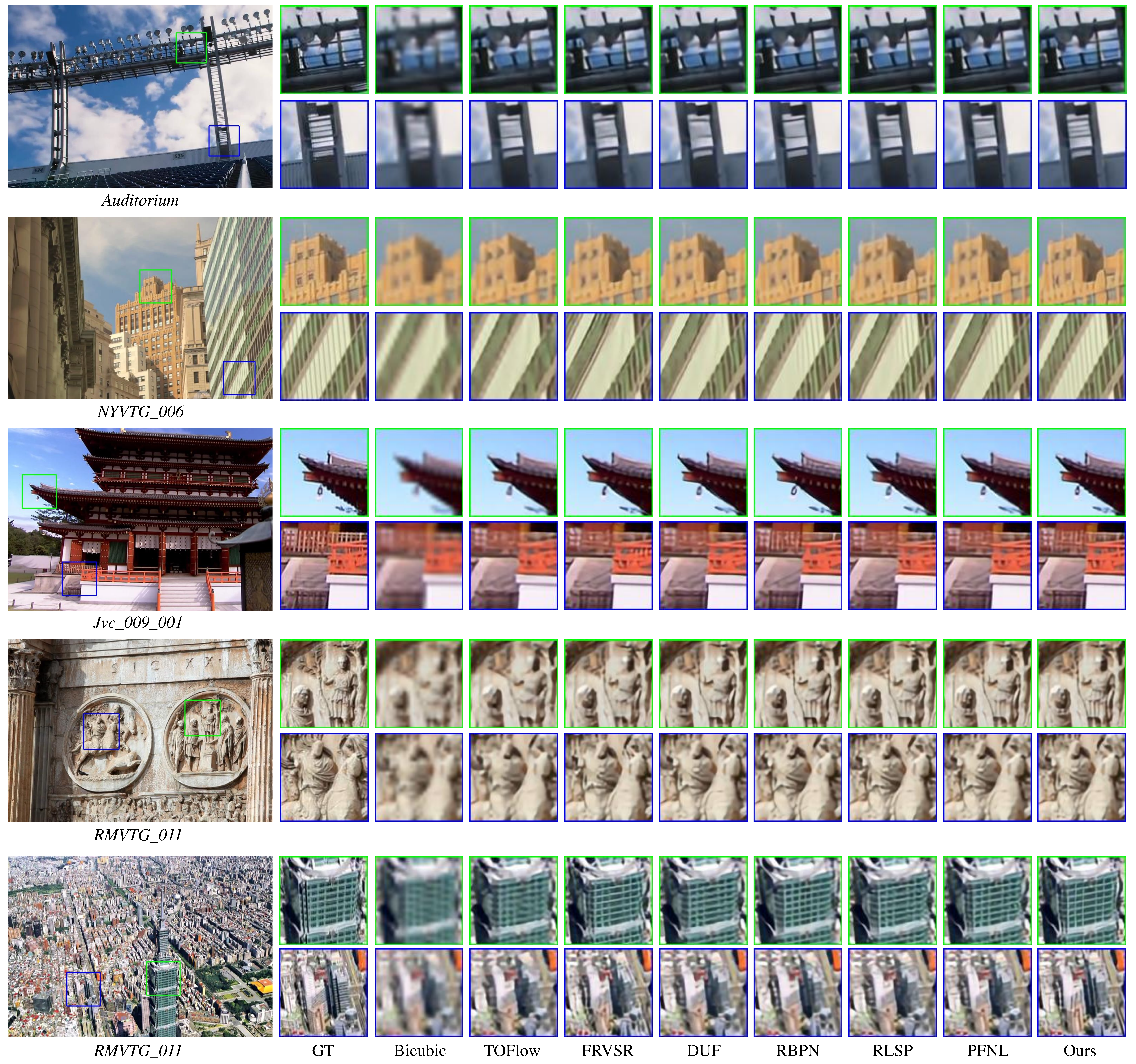}
		\vspace{-5mm}
		\caption{\small Qualitative comparison on the \textbf{UDM10}~\cite{yi2019progressive} and \textbf{SPMCS}~\cite{tao2017detail} datasets for 4$\times$ VSR. Zoom in for better visualization.}
		\label{soat} 
		\vspace{-3mm}
	\end{figure*}

	\begin{figure*}[t]
		\centering
		\includegraphics[width=0.95\textwidth] {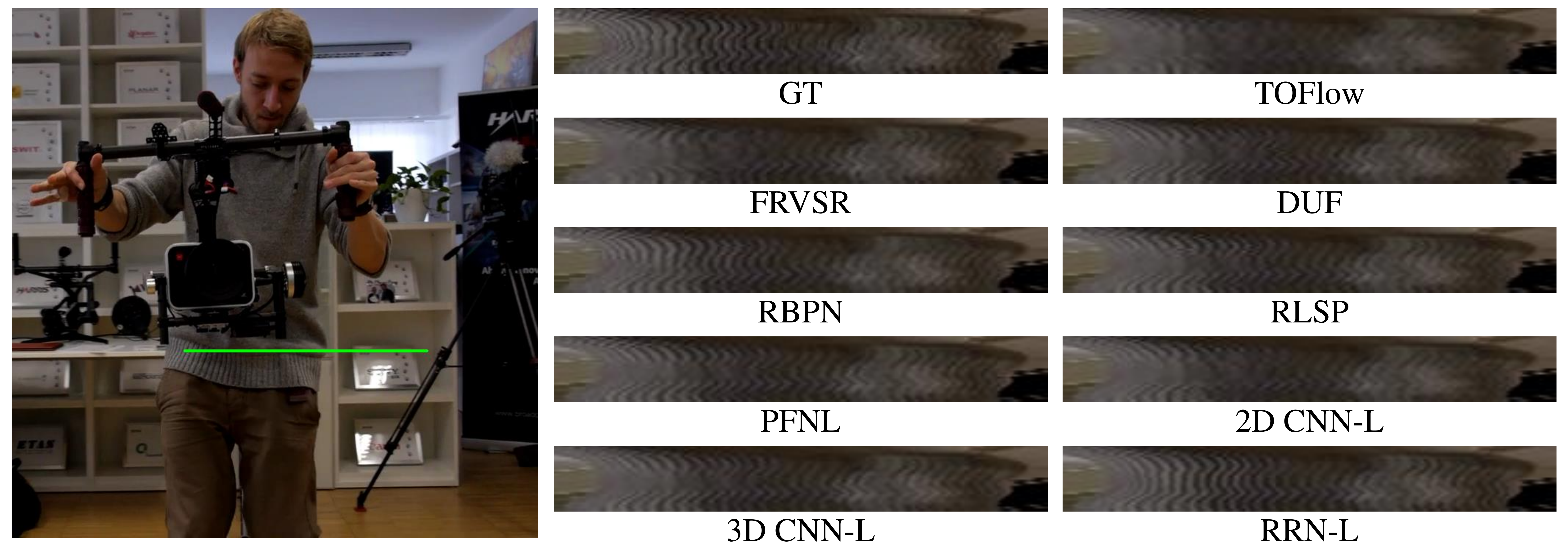}
		\vspace{-2mm}
		\caption{\small Visualization of temporal consistency for the \textit{Photography} sequence in UDM10~\cite{yi2019progressive}. The temporal profile is produced by recording a single-pixel line (\textbf{\color{green}green} line) spanning time and stacked vertically.}
		\label{profile} 
		\vspace{-3mm}
	\end{figure*}
	
	\subsection{Comparison with different temporal modeling methods}
	
	In this part, we compare three temporal modeling approaches, including 2D CNN, 3D CNN and RNN on Vid4~\cite{liu2013bayesian}, SPMCS~\cite{tao2017detail} and UMD10~\cite{yi2019progressive} datasets. The quantitative and qualitative results are shown in Tab.~\ref{model_tab} and Fig.~\ref{modeling}, respectively. We also present the trade-off between runtime and accuracy in Tab.~\ref{model_tab}. 
	\\

	In CNN-based methods, 3D CNN-S and 3D CNN-L outperform 2D CNN-S and 2D CNN-L by a large margin. However, VSR with 3D CNN is very time-consuming as shown in Tab.~\ref{model_tab}, where 3D CNN-L is almost ten times slower than 2D CNN-L on processing an LR frame of size $320\times 180$. Comparing with CNN based methods, RNN is highly computational efficiency and achieves excellent results with fewer parameters. RRN-L is 0.44, 0.20 and 0.54 dB higher than 3D CNN-L on Vid4, SPMCS and UDM10, respectively, and meanwhile being more than $23\times$ faster. Moreover, RNN-S and RNN-L can produce a 720p video sequences in 33fps and 22fps, respectively. The qualitative results also show that RRN-L can produce finer details and fewer artifacts than 2D CNN-L and 3D CNN-L. In addition, we visualize the temporal profiles in Fig.~\ref{profile}. RRN-L produces temporal consistent frames and suffers less flickering artifacts than other temporal modeling methods. 
	\\
	To investigate the information flow of different temporal modeling methods for handling a long video sequence, we plot the PSNR over time for $Calendar$ sequence in Vid4. As shown in Fig.~\ref{modeling} (b),the RNN-based method falls behind the CNN-based methods at the first five frames. By accumulating information over time, the RNN-based method outperforms CNN-based methods from the fifth frame. More interestingly, the CNN-based methods suffer from performance degradation at the fifth frames, but the RNN-based method keeps improving, which demonstrates that the information accumulation in the previous hidden state provides a complementary information for recovering missing details.

	\textbf{Necessity of residual connection in the hidden state of RNN.}
	To investigate the necessity of residual learning in the hidden state, we create a simple baseline by removing by simply stacking the convolutional layer in a hidden state. The advanced model is obtained by incorporating the identity mapping in a hidden state. The PSNR (dB) and SSIM results on Vid4 are shown in Tab.~\ref{blocks}. The quantitative results are measured on the luminance (Y) channel. As shown in Tab.~\ref{blocks}, the best performance of the baseline model is 27.09 dB in PSNR, which uses three blocks as the hidden state. However, it suffers from gradient vanishing when increasing the number of blocks to the number of four. With the help of a residual connection in the hidden state, stable improvement is achieved when increasing the number of blocks. These results illustrate that identity mapping can not only stabilize training but also boost the VSR performance. Note that the performance of RRN can be further improved by adopting more blocks.
	
	\begin{table}[h]
		\centering
		\scalebox{0.52}
		{	
			\begin{tabular}{l||c|c|c|c|c|c|c|c|c}
				%\begin{tabular}{llllllllll}
				%\resizebox{\textwidth}{12mm}{   
				\hline   \hline  	
				Blocks &2 &3  &4  &5  &6   	&7  &8 &9 &10
				\\ \hline	
				RRN w/o residual    &26.70/0.8050&26.75/0.8108 &$\ast$   &$\ast$   &$\ast$  &$\ast$  &$\ast$ &$\ast$  &$\ast$
				
				\\ \hline
				RRN w/ residual  &26.97/0.8119  &27.09/0.8286 &27.20/0.8334   &27.38/0.8385 &27.42/0.8434 &27.53/0.8433 & 27.65/0.8468 & 27.67/0.8476 &27.69/0.8488
				\\ \hline
			\end{tabular}
		}
		\vspace{3mm}
		\caption{\small Ablation on the residual learning in the hidden state of RRN. ``$\ast$'' represents that the model in training suffers from gradient vanishing. }
		\vspace{-3mm}
		\label{blocks}
	\end{table}

	\begin{table}[t]
		\centering
		\scalebox{0.46}
		{	
			\begin{tabular}{l||c|c|c|c|c|c|c|c|c|c}
				%\begin{tabular}{llllllllll}
				%\resizebox{\textwidth}{12mm}{   
				\hline   \hline  	
				Method &Bicubic &SPMC$^\dagger$~\cite{tao2017detail}  &TOFLOW~\cite{xue2019video} 
				&FRVSR~\cite{sajjadi2018frame} &DUF~\cite{jo2018deep}  &RBPN~\cite{haris2019recurrent}
				&EDVR~\cite{wang2019edvr}
				&RLSP~\cite{Fuoli-arxiv19-rlsp}
				&PFNL~\cite{yi2019progressive}
				&RRN-L (Ours)
				\\	
				\#~Param. [M] &N/A &-   &1.4 &5.1 &5.8 &12.8 &20.1 &4.3 &3.0 &3.4\\
				Runtime [ms] &N/A &-  &1658 &129 &1393 &3482 &621 &50 &295 &45 \\ \hline \hline
				
				Vid4 (Y) &21.80/0.5426 &25.52/0.76  &25.85/0.7659 &26.48/0.8104   &27.38/0.8329  &27.17/0.8205 &27.35/0.8264 &{\color{blue}27.48}/{\color{blue}0.8388} &27.16/0.8365 &{\color{red}27.69}/{\color{red}0.8488}
				\\ 
				Vid4 (RGB) &20.37/0.5106 &-/-   &24.39/0.7438 &25.01/0.7917 &{\color{blue}25.91}/0.8166 &25.65/0.7997 &25.83/0.8077   &25.69/0.8153 &25.67/{\color{blue}0.8189} &{\color{red}26.16}/{\color{red}0.8209} \\ \hline \hline

				SPMCS (Y) &23.29/0.6385 &-/- &27.86/0.8237 &28.16/0.8421 &29.63/0.8719  &29.73/0.8663 &-/-   &29.59/0.8762 &{\color{blue}29.74}/{\color{blue}0.8792} &{\color{red}29.84}/{\color{red}0.8827}
				\\
				
				SPMCS (RGB)) &21.83/0.6133 &-/- &26.38/0.8072 &26.68/0.8271 &28.10/{\color{blue}0.8582} &{\color{blue}28.23}/0.8561  &-/- &27.25/0.8495 &27.24/0.8495  &{\color{red}28.28}/{\color{red}0.8690}	\\ \hline \hline

				UDM10 (Y) &28.47/0.8523   &-/-  &36.26/0.9438 &37.09/0.9522 &38.48/0.9605  &38.66/0.9596   & -/- &38.48/0.9606 &{\color{blue}38.74}/{\color{blue}0.9627} &{\color{red}38.96}/{\color{red}0.9644}
				\\ 
				UDM10 (RGB) &27.05/0.8267 &-/-   &34.46/0.9298  &35.39/0.9403  &36.78/0.9514  &36.53/0.9462   &-/-  &36.39/0.9465 &{\color{blue}36.91}/{\color{blue}0.9526} &{\color{red}37.03}/{\color{red}0.9534} \\ \hline
			\end{tabular}
		}
		\vspace{3mm}
		\caption{\small Quantitative comparison (PSNR(dB) and SSIM) on \textbf{Vid4}~\cite{liu2013bayesian}, \textbf{SPMCS}~\cite{tao2017detail} and \textbf{UDM10}~\cite{yi2019progressive} for $4\times$VSR, respectively. `$\dagger$' means the values are taken from original publications or calculated by provided models. Y and RGB indicate the evaluation on luminance channel or RGB channels, respectively. Runtime is calculated on an LR image of size 320$\times$180. {\color{red}Red} text indicates the best and {\color{blue} blue} text indicates the second best performance.Best view in color.}
		\vspace{-3mm}
		\label{soat_tabel}
	\end{table}
	\subsection{Comparison with state-of-the-art methods}
	We compare our best model (RRN-L) with eight state-of-the-art VSR approaches: SPMC~\cite{tao2017detail}, TOFlow~\cite{xue2019video}, FRVSR~\cite{sajjadi2018frame}, DUF~\cite{jo2018deep}, RBPN~\cite{haris2019recurrent}, EDVR~\cite{wang2019edvr}, RLSP~\cite{Fuoli-arxiv19-rlsp} and PFNL~\cite{yi2019progressive}.
	%	The first three methods adopt explicit motion estimation and compensation, and the other methods utilizes motion information in an implicitly manner.
	SPMC, TOFlow and FRVSR apply for explicit motion estimation and compensation. EDVR conducts motion alignment in feature level. RBPN also computes optical flow but uses it as additional input instead of explicit motion compensation. DUF and PFNL use an advanced temporal integration network to utilize motion information in an implicit way. RLSP is the most related work, which also propagates historical information in feature space. However, the design of the hidden state of RLSP is simple, which easily causes gradient vanishing problem (see Tab~\ref{blocks}). Most of the previous methods use different training sets and different down-sampling operations. For fair comparison, we fix the down-sampling filters, $\textit{i.e.} \sigma=1.6$, and carefully re-implement these methods on the public Vimeo-90k dataset. The quantitative results on Vid4, SPMCS and UDM10 are shown in Tab.~\ref{soat_tabel}. We can see that methods with explicit motion compensation do not perform very well. By carefully analyzing, the occlusion or complex motion easily influences the per-pixel motion estimation, such as optical flow. Inaccurate motion estimation would introduce artifacts which deteriorate super-resolution performance. As shown in Tab.~\ref{soat_tabel}, our method outperforms the CNN-based methods~\cite{xue2019video,tao2017detail,jo2018deep,haris2019recurrent,wang2019edvr,yi2019progressive} by a large margin, and meanwhile runs been $70\times$ and  $6\times$ faster than the recent proposed RBPN and PFNL, respectively. Comparing with other RNN-based methods~\cite{sajjadi2018frame,Fuoli-arxiv19-rlsp}, our method outperforms FRVSR by a large margin even with fewer parameters and $2\times$ faster. Our method achieves comparable runtimes with RLSP, but it outperforms RLSP by 0.21, 0.25 and 0.48 dB in PSNR on Vid4, SPMCS and UDM10, respectively.
	Qualitative comparisons are presented in Fig.~\ref{soat}. The proposed method produces sharper edges and finer details than other VSR methods. In addition, our method produces temporal-consistent results than the previous methods, as shown in Fig.~\ref{profile}

	\section{Conclusion}
	Video super-resolution is an important task, which has drawn much attention in both research and industrial communities. We comprehensively investigate and compare three commonly used temporal modeling methods for video super-resolution, including 2D CNN with early fusion, 3D CNN with slow fusion and RNN. For a fair comparison, all models are developed on the public Vimeo-90k dataset with the fixed down-sampling filters and loss function. Extensive experiments on Vid4, SPMCS and UDM10 benchmarks, demonstrate RNN is highly efficient and benefit in dealing with the VSR problem. In addition, we also propose a novel hidden state structure for recurrent network, termed as RRN. The proposed method achieves state-of-the-art performance on three benchmarks. 
	\section*{Acknowledgement}
	This work was supported in part by the National Natural Science Foundation of China under Grant 61701277 and Grant 61771288 and in part by the State Key Development Program in 13th Five-Year under Grant 2017YFC0821601.
	\bibliography{egbib}
\end{document}